\magnification\magstep 1
\vsize=23 true cm
\topinsert\vskip 1 true cm
\endinsert
{\centerline{\bf{SELF-DUALITY IN MAXWELL-CHERN-SIMONS THEORIES WITH}}}
{\centerline{\bf{NON MINIMAL COUPLING WITH MATTER FIELD}}}
\vskip 2 true cm
{\centerline{\bf{Fr\'ed\'eric Chandelier$^a$,Yvon Georgelin$^a$, Thierry Masson$^b$ and 
Jean-Christophe Wallet$^a$}}}
\vskip 2 true cm
{\centerline{$^a$Groupe de Physique Th\'eorique, Institut de Physique
Nucl\'eaire}}
{\centerline{F-91406 ORSAY Cedex, France}}
\vskip 1 true cm
{\centerline{$^b$Laboratoire de Physique Th\'eorique (U.M.R. 8627), 
Universit\'e de Paris-Sud}}
{\centerline{B\^at. 211, F-91405 ORSAY Cedex, France}}
\vskip 2 true cm
{\bf{Abstract:}} We consider a general class of non-local MCS models whose
usual minimal coupling to a conserved current is supplemented with a
(non-minimal) magnetic Pauli-type coupling. We find that the considered models
exhibit a self-duality whenever the magnetic coupling constant reaches a
special value: the partition function is invariant
under a set of transformations among the parameter space (the duality
transformations) while the original action and its dual counterpart have the
same form. The duality transformations have a structure similar to the one
underlying self-duality of the (2+1)-dimensional $Z_n$-abelian Higgs model with Chern-Simons
and bare mass term.\par

\vskip 4 true cm
(November 2000 )
\vskip 0.5 true cm
LPT-00/121
\vfill\eject
Infinite discrete symmetries, duality symmetries have received a continuous
attention within supersymmetric gauge theories [1], string theories [2],
extended sine-Gordon models [3], and statistical systems [4] together with
condensed matter models applied for instance to quasi-planar high-T$_C$
materials and Josephson junction arrays [5] and Quantum Hall Effect (QHE) [6].
Recall that basically a duality symmetry occurs within a model when there
exits a set of transformations in the corresponding parameter space (the
space of the coupling constants) that leaves the partition function
invariant while the original action is mapped into an action having the same
form whose coupling constants transform according to the duality symmetry.
Such a model is called a self-dual model. The occurence of a duality
symmetry within a model may well have interesting consequences since it can
be used to derive (exact) non perturbative results and/or to obtain
information on the corresponding phase diagram, as it has been done in [4]
where the relevant duality symmetry has been related to the modular group
which, in that context, can be viewed as a generelization of the old $Z_2$
Kramers-Wannier duality of the 2-dimensional Ising model.\hfill\break
Notice that modular 
symmetries provide an appealing way to understand (most
of) the observed features of the QHE [7] and may well account for the apparently
observed superuniversality occuring among the transitions [8]. In
particular, two subgroups of the modular group have been recently singled
out [9,10] as potential candidates for a symmetry underlying the QHE and have been
used successfully to derive a model for the classification of the Hall
states as well as to built a global phase diagram stemming from a physically
admissible $\beta$-function reproducing nicely several observed properties
of the QHE [9,10] (such as stability of the Hall plateaus, the semi-circle law, and
the (qualitative shape of) crossovers). \hfill\break
As far as specific models are 
concerned, self-duality has been shown to
occur within the (2+1)-dimensional abelian Higgs model with Chern-Simons
(CS) term and bare mass term (on the lattice) [11] and its relevance to some
condensed matter systems has been examined [11]. More recently, a general
class of non-local Maxwell-Chern-Simons (MCS) models (on the lattice) has
been considered [12] and shown to exhibit self-duality in the low energy
long-wavelength limit which has been further used in an attempt to explain
some global properties of the transitions between plateaus [12].\hfill\break
In this letter, we 
consider a general class of non-local MCS models whose
usual minimal coupling to a conserved current is supplemented with a
(non-minimal) magnetic Pauli-type coupling {\footnote *{\sevenrm{Notice that such models may be viewed as
effective models in the same way as those considered in [12] so that non local
and/or non renormalizable terms can be involved in the corresponding
action.}}}. We find that, for a unique value of the magnetic coupling constant, the considered models
exhibit a self-duality: the partition function is invariant
under a set of transformations among the parameter space (the duality
transformations) while the original action and its dual counterpart have the
same form. The duality transformations have a structure similar to the one
exhibited in [11] (although the parameters involved in both case are
different).\hfill\break
The general class of MCS models we consider 
is {\it{defined}} by the following non-local $U(1)$-gauge invariant
action{\footnote\dag{\sevenrm{We choose $\hbar$$=c$$=1$, $g_{\mu\nu}$=diag(+1,-1,-1),
$\epsilon_{012}$=+1, $dp\equiv{{d^3p}\over{(2\pi)^3}}$ and $p$ is the
momentum; $\int_{x,y,p}\equiv\int dxdydp$}}}:
$$S=\int_{x,y,p}e^{ip(x-y)}\big({{\sigma_L}\over{{\sqrt{p^2}}}}
{{1}\over{4}}f_{(x)\mu\nu}f_{(y)}^{\mu\nu}+{{1}\over{4}}\theta\epsilon_{\mu\nu\rho}
a_{(x)}^\mu f_{(y)}^{\nu\rho}
+a_{(x)\mu}J_{(y)}^{\mu}+{{1}\over{2}}{{\kappa}\over{{\sqrt{p^2}}}}\epsilon_{\mu\nu\rho}f_{(x)}^{\mu\nu}
J_{(y)}^\rho\big) \eqno(1),$$
where $f_{\mu\nu}=\partial_\mu a_\nu-\partial_\nu a_\mu$ is the field
strength for the gauge potentiel $a_\mu$, $J_\mu$ is a ($U(1)$-gauge
invariant) conserved current ($\partial_\mu J^\mu=0$) and $\sigma_L$, $\theta$ and $\kappa$ 
are constant. In (1), the first two terms represent the MCS
part of the action whose Maxwell part is assumed to be non-local while the
Chern-Simons part, that is, the term involving $\theta$, remains local. Notice that
similar non-local Maxwell terms (with such a $1/{\sqrt{p^2}}$ dependance)
have already been introduced in some earlier works to describe the
coupling of a $(3+1)$-dimensional electromagnetic field to systems
involving particles and currents confined to a plane [13]. Moreover, the class
of models considered more recently in [12] in an attempt to describe some
features of the physics of the transition between Hall states has been shown
to exhibit a rich underlying symmetry structure {\footnote\ddag{\sevenrm{In
the long wavelength limit}}} whenever the Maxwell term
has a non local behavior similar to the one considered in (1). Notice
finally that the constant $\sigma_L$ in (1) can be naturally interpreted in
a condensed matter framework as a longitudinal conductivity.\par
The last two terms in the action (1) describe the coupling 
of the gauge potential $a_\mu$ to the conserved current $J^\mu$. 
Observe that in (1) the usual minimal coupling term (4st term) is 
supplemented by an additional non-local magnetic Pauli-type coupling whose 
strength is given by $\kappa$. A 
non-minimal magnetic Pauli-type coupling occuring beyond the usual minimal 
coupling has been already considered from somehow different 
viewpoints [14,15]. In particular, it has been shown that the MCS
theory (in its usual {\it{local}} version) non minimally coupled to 
matter exhibits an interesting property whenever the strenght of the 
magnetic coupling reaches a special value [14,15]. In particular, at this value, one obtains an
anyonic behavior similar to the one usually obtained within a pure
Chern-Simons theory minimally coupled with matter [14]. In the present
situation, the magnetic coupling term in the action is required to be
non-local with a momentum dependance similar to the one for the Maxwell
term.\par
The action (1) can be put into a more convenient form by using the
well-known fact that a conserved current in (2+1)-dimension can be written
{\it{locally}} as the curl of a (pseudo)vector field. Namely, one has
$J^\mu=\epsilon_{\mu\nu\rho}\partial^\nu
v^\rho\equiv{{1}\over{2}}\epsilon_{\mu\nu\rho}w^{\nu\rho}$ where
$w_{\mu\nu}=\partial_\mu v_\nu-\partial_\nu v_\mu$ 
from which (1) can be rewritten as
$$S=\int_{x,y,p}e^{ip(x-y)}\big[{{1}\over{4}}{{\sigma_L}\over{{\sqrt{p^2}}}}
f_{(x)\mu\nu}f_{(y)}^{\mu\nu}+{{1}\over{4}}\theta\epsilon_{\mu\nu\rho}
a_{(x)}^\mu f_{(y)}^{\nu\rho}\big]
+\big[{{1}\over{2}}\epsilon_{\mu\nu\rho}a_{(x)}^\mu
w_{(y)}^{\nu\rho}
+{{1}\over{2}}{{\kappa}\over{{\sqrt{p^2}}}}f_{(x)\mu\nu}
w_{(y)}^{\mu\nu}\big] \eqno(2).$$
In the following, the first (resp. second) term into brackets in (2) will be
denoted by $S_1(\sigma_L,\theta;a)$ (resp. $S_2(\kappa;v)$ or equivalently
$S_2(\kappa;J)$).
It can be easily realized that the mixed Chern-Simons and mixed Maxwell term
occuring in (2) render its structure somehow similar to the one 
for some planar gauge theories that
have been already considered as being of possible relevance for the
description of 2-dimensional condensed matter systems such as Josephson
junction arrays and quasi-planar high-T$_C$ materials [5], in which the
current is often assumed to describe the behavior of matter degrees of
freedom. Notice that (2) is now invariant under an additional $U(1)$-gauge
symmetry stemming from the introduction of the field $v_\mu$.\par
Let us present now in details the main result of this letter. We find that, for a
unique value of the magnetic coupling constant
$\kappa$, the considered model
exhibits a self-duality. By self-dual model, we mean that the original model
and its dual (obtained through duality transformations to be precised in a
while) have the same form. More precisely, the self-duality occuring for the
pure MCS part of the action (2), which is spoiled when a minimal
coupling to a conserved current is introduced (as it can be expected), is
recovered by the addition of a non-minimal magnetic coupling term whose
magnetic coupling constant
$\kappa$ takes a special value given by 
$$\kappa=\kappa_c\equiv{{\sigma_L}\over{\theta}} \eqno(3).$$\par
This statement reflects itself into the partition function {\footnote\dag{\sevenrm{from now the 
gauge fixing is implicitely
assumed. Furthermore suitable integration over space and/or momentum in the argument of the
exponential is understood for the sake of brievity as well as the exp ip(x-y) factor
which is not explicitely written but can be reinstalled trivially.}}}built from (2) as
follows:${\cal{Z}}= \int[{\cal{D}}a]\exp
i(S_1(\sigma_L,\theta;a)+S_2(\kappa=\kappa_c;v))$$\equiv$ 
$$\int[{\cal{D}}a]\ \exp i\big(\int{{1}\over{4}}{{\sigma_L}\over{{\sqrt{p^2}}}}
f_{(x)\mu\nu}f_{(y)}^{\mu\nu}+{{1}\over{4}}\theta\epsilon_{\mu\nu\rho}
a_{(x)}^\mu f_{(y)}^{\nu\rho}
+{{1}\over{2}}\epsilon_{\mu\nu\rho}a_{(x)\mu}w_{(y)}^{\nu\rho}
+{{1}\over{2}}{{\sigma_L}\over{{\theta\sqrt{p^2}}}}f_{(x)\mu\nu}
w_{(y)}^{\mu\nu}\big) \eqno(4)$$
$$=\int[{\cal{D}}a]\exp i\big(\int
{{\sigma_L}\over{4(\sigma_L^2+\theta^2){\sqrt{p^2}}}}
f_{(x)\mu\nu}f_{(y)}^{\mu\nu}+{{\theta}\over{4(\sigma_L^2+\theta^2)}}\epsilon_{\mu\nu\rho}
a_{(x)}^\mu f_{(y)}^{\nu\rho}(y)$$
$$+{{1}\over{2(\sigma_L^2+\theta^2)^{1/2}}}\epsilon_{\mu\nu\rho}a_{(x)\mu}w_{(y)}^{\nu\rho}
+{{\sigma_L}\over{{2\theta(\sigma_L^2+\theta^2)^{1/2}\sqrt{p^2}}}}f_{(x)\mu\nu}
w_{(y)}^{\mu\nu}\big) \eqno(5),$$
which, by further introducing the dual current $J^D_\mu$ defined by
$J^D_\mu=(\sigma_L^2+\theta^2)^{-1/2}J_\mu$
can finally be cast into the form
$$Eqn.(5)= \int[{\cal{D}}a]\
\exp\big(i(S_1({{\sigma_L}\over{(\sigma_L^2+\theta^2)}},
{{\theta}\over{(\sigma_L^2+\theta^2)}};a)+S_2(\kappa=\kappa_c;J^D))\big)
\eqno(6).$$
From a comparison of (4) and (6), it can be easily realized that the model
under consideration is self-dual under the duality transformation given by
$$\sigma_L \to {{\sigma_L}\over{(\sigma_L^2+\theta^2)}}\equiv\sigma_L^D\ ;\ \theta \to
{{\theta}\over{(\sigma_L^2+\theta^2)}}\equiv\theta^D\ {\hbox{ or
equivalently}}\ z\to {{1}\over{\bar z}} {\hbox{with}}\ z=\sigma_L+i\theta
\eqno(7a;b;c;d),$$
where in (7c) $\bar z$ denotes the complex conjugate of $z$.\par
Let us now derive of the result of this letter. From now on,
$\kappa=\kappa_c$. We will present two ways of
obtaining this result. A possible and standard way to explore the potential 
occurence of self-duality properties
within a model is to apply the machinery of the Hubbard-Stratonovitch (HS)
transformation to the corresponding partition function. Recall that the HS
transformation is based on the following functional relation which can be
generically written as{\footnote\ddag{\sevenrm{for real valued $\Phi$}}}
$$e^{i\int
dx{{1}\over{2}}\alpha\Phi(x)^2}=\int[{\cal{D}}\Lambda]e^{{{-i}\over{2}}\int
dx {{1}\over{\alpha}}\Lambda(x)^2+\Lambda(x)\Phi(x)} \eqno(8),$$
where $\alpha$ is some constant and $\Lambda$ is an auxilliary field (the HS
field) associated to $\Phi$. The HS transformation can be applied by brute
force to the partition function built from (2). The corresponding derivation
is standard but cumbersome and we will now summarize its main steps. First, the action (2) 
can be put into a quadratric form which can be written
in momentum space as
$$S=\int dp\ zA_+^\mu(-p)A_{+\mu}(p)+\bar z
A_-^\mu(-p)A_{-\mu}(p)+{{i2}\over{z-\bar z}}\big(zA_+^\mu(-p)V_{+\mu}(p)+\bar z
A_-^\mu(-p)V_{-\mu}(p)\big)
\eqno(9),$$
where we have introduced for convenience the complex coupling constant
$z=\sigma_L+i\theta$ and $A_\pm^\mu(p)=(u_1T^{\mu\nu}(p)\pm
u_2C^{\mu\nu}(p))a_\nu(p)$ in which the real numbers $u_1$ and $u_2$ verify
$u_1u_2=({\sqrt{p^2}})/8$, $u_1^2-u_2^2=({\sqrt{p^2}})/4$
and the parity conserving $T_{\mu\nu}(p)$ and parity violating operator $C_{\mu\nu}(p)$
are defined by
$$T_{\mu\nu}(p)=g_{\mu\nu}-{{p_\mu p_\nu}\over{p^2}}\ ;\
C_{\mu\nu}(p)=\epsilon_{\mu\nu\rho}{{p^\rho}\over{{\sqrt{p^2}}}}
\eqno(10a;b),$$
while $V_\pm(p)$ can be easily read off from (9) and (2) combined with the
expressions for $A_\pm(p)$. Next, to perform the HS transformation, one has to adapt (8) to the present
situation. This gives rise to
$$e^{i\int dpzA_+^\mu(-p)A_{+\mu}(p)}=\int [{\cal{D}}\Lambda_+]e^{-i\int dp
{{1}\over{z}}\Lambda^\mu_+(-p)\Lambda_{+\mu}(p)-\Lambda^\mu_+(-p)A_{+\mu}(p)-
A^\mu_+(-p)\Lambda_{+\mu }(p)}  \eqno(11),$$
where $\Lambda_+$ is the HS field for $A_+$ together with a similar
expression relating $A_-$ to its HS partner deduced from (11) through the
following substitution: $\Lambda_+\to\Lambda_-$, $A_+\to A_-$, $z\to \bar
z$. Inserting these relations into the partition function built from (9)
and further integrating out $a_\mu$, one obtains a constraint given by
$$\big(u_1T^{\mu\nu}+u_2C^{\mu\nu}\big)\big(-\Lambda_{+\nu}+{{iz}\over{z-\bar
z}}V_{+\nu}\big)=
\big(u_1T^{\mu\nu}-u_2C^{\mu\nu}\big)\big(\Lambda_{-\nu}-{{i\bar
z}\over{z-\bar z}}V_{-\nu}\big)  \eqno(12),$$
stemming from the terms linear in $a_\mu$ appearing in the action. This constraint 
is then found to be solved by setting
$$\Lambda^\mu_+=-{{1}\over{2}}\big(u_2T^{\mu\nu}+
u_2C^{\mu\nu}\big)\tilde a_\nu+{{iz}\over{z-\bar z}}V_+\ ;\ 
\Lambda^\mu_-=-{{1}\over{2}}\big(u_2T^{\mu\nu}-
u_2C^{\mu\nu}\big)\tilde a_\nu+{{i\bar z}\over{z-\bar z}}V_-
\eqno(13a;b),$$
where $\tilde a_\mu$ is some vector field. These relations permit one to
reexpress the resulting partition function as a functional integral over
$\tilde a_\mu$ while the corresponding action involved in it takes the form
$$S={{1}\over{2}}\int dp\big({{1}\over{\vert z\vert^2}}\tilde 
a^\mu(-p)K_{\mu\nu}(p)\tilde a^\nu(p)+{{1}\over{(z-\bar
z)^2}}v^\mu(-p)K_{\mu\nu}(p)v^\nu(p\big))
\eqno(14),$$
where $K_{\mu\nu}(p)={\sqrt{p^2}}(\sigma_LT_{\mu\nu}(p)+i\theta
C_{\mu\nu}(p))$. Now
the last term in (14) represents a current-current interaction which
can be reabsorbed through the following field redefinition:\hfill\break
$\tilde
a_\mu=a_\mu^\prime+i{{\vert z\vert}\over{z-\bar z}}v_\mu$. This, combined
with the partition function and further introducing the dual current
$J_\mu^D$ defined previously gives rise to (6).\par
The possibility of reabsorbing the current-current term through the above
mentionned field
redefinition comes out from the fact that the action (2) can be already
splitted into two independent MCS parts only when $\kappa=\kappa_c$ by introducing a similar field
redefinition. This property has already been noticed in earlier works [14,15].
Indeed, define $a_\mu=b_\mu-{{1}\over{2\theta}}v_\mu$ which, reported into (2)
{\it{and}} setting $\kappa=\kappa_c$, yields $S\vert_{\kappa=\kappa_c}\equiv S(z,\bar z;a,v)
=S^{MCS}(z,\bar z;b)+S^{MCS}(z,\bar z;v)$ in which 
$$S^{MCS}(z,\bar z;b)=\int dp
{{\sigma_L}\over{4{\sqrt{p^2}}}}B_{\mu\nu}(-p)B^{\mu\nu}(p)+{{1}\over{4}}\theta
\epsilon_{\mu\nu\rho}b^\mu(-p)B^{\nu\rho}(p) \eqno(15a),$$
$$S^{MCS}(z,\bar z;v)=\int dp-{{1}\over{4}}{{\sigma_L}\over{\theta^2{\sqrt{p^2}}}}
w_{\mu\nu}(-p)w^{\mu\nu}(p)-{{1}\over{4\theta}}\epsilon_{\mu\nu\rho}v^\mu(-p)w^{\nu\rho}(p)\eqno(15b),$$
where $B_{\mu\nu}$ denotes the field strenght for $b_\mu$. The
combination of this last property together with the fact that the pure MCS
part of (2) is self-dual under the duality transformation (7) permits one to
obtain an alternative and simpler way to show that the model (2) is
self-dual under (7) when $\kappa=\kappa_c$. Indeed, the corresponding
partition function can be written as
$${\cal{Z}}(z,\bar z;v)\equiv\int[{\cal{D}}a]e^{iS(z,\bar
z;a,v)}=\int[{\cal{D}}b]e^{iS^{MCS}(z,\bar
z;b)}\times e^{iS^{MCS}(z,\bar z;v)}  \eqno(16),$$
where $S^{MCS}(z,\bar z;b)$ and $S^{MCS}(z,\bar z;v)$ are given by (15a) and
(15b) and the second equality is obtained through the field redefinition
$a_\mu=b_\mu-{{1}\over{\theta}}v_\mu$. Now, it is easy to prove that $S^{MCS}(z,\bar z;b)$
is self-dual under (7). The corresponding derivation is also based on a
suitable application of the HS transformation and can be obtain as a simple
byproduct of the previous derivation by setting $v_\mu=0$ in all. From this
latter self-duality property, it follows that
$${\cal{Z}}(z,\bar z;v)=\int[{\cal{D}}\tilde b]e^{iS^{MCS}({{1}\over{z}},{{1}\over{\bar
z}};\tilde b)}\times e^{iS^{MCS}(z,\bar z;v)}  \eqno(17).$$
Setting now $\tilde b_\mu=\tilde a_\mu+{{1}\over{\theta^D}}v^D_\mu$ in (17)
where $\theta^D$ is still given by (7b), it can be easily seen that the
terms involving only $v_\mu$ in the action appearing in (17) cancel each
other {\it{provided}} $v_\mu^D=v_\mu/{\sqrt{z\bar z}}$, which is consistent with
the definition of the dual current introduced above and duality. Therefore, one 
obtains ${\cal{Z}}(z,\bar z;v)={\cal{Z}}
{\hbox{eqn.(6)}}={\cal{Z}}({{1}\over{z}},{{1}\over{\bar z}};v^D)$, 
showing that (2) evaluated at $\kappa=\kappa_c$ is self-dual under (7).\par
Let us now discuss and comment the above result. First, we point out that
the introduction of a non-minimal Pauli-type magnetic coupling whose
coupling constant takes a unique value given by (3) is essential to render the model
self-dual. Had we only coupled the current to the MCS part
in a minimal way, then self-duality of the resulting model would have been
sploit through the appearence in the dual model (obtained from the
HS transformation) of current-current interaction terms
($\sim (z+\bar z)J^\mu J_\mu$). The introduction of an arbitrary magnetic
coupling does not improve the situation since it generates other unwanted
current-current terms so that the original action and its dual counterpart
do not have the same form except at the self-dual point $z=-\bar z$ where
all those unwanted terms vanish. However, when $\kappa=\kappa_c$, the dual
action and the original one have the same form. More precisely, the
current-current terms can be combined in a way that a suitable field
redefinition allows one to put the dual action into a form similar to the
one for the original action. This reflects the fact that in the original
action the $v_\mu$ part can be decoupled from the $a_\mu$ part through a
field redefinition which is not altered by the "dualization" procedure and
can be used to combine the current-current terms in a way that the form of
the original action is preserved at the level of the dual action (which of
course involves a similar decoupling property).\par 
Next, we observe that similar duality transformations have been exhibited
within the $Z_n$-abelian Higgs model with Chern-Simons
term and bare mass term on the lattice{\footnote\dag{\sevenrm{where now the corresponding complex coupling
constant depends on the mass parameter of the bare mass term involved in
this model}}} [11]. As a side remark, we notice that when $\sigma_L=0$ for
which only the Chern-Simons term (2) survives (while the magnetic term
vanishes), the transformation (7) reduces to $\theta\to 1/\theta$ which now
corresponds to a $Z_2$ Kramers-Wannier duality. \par
It is instructive to compute from (4) the effective action, obtained by
integrating out the $a_\mu$ field. After some calculations, we find that this
latter is given by
$$S_{eff}=\int_{x,y,p}e^{ip(x-y)} \big(-{{\sigma_L}\over{4\theta^2{\sqrt{p^2}}}}
w_{(x)\mu\nu}w_{(y)}^{\mu\nu}-{{1}\over{4\theta}}\epsilon_{\mu\nu\rho}v_{(x)}^\mu
w_{(y)}^{\nu\rho}\big)
\eqno(18),$$
where the first term corresponds to a (non local) current-current
interaction for $J_\mu$ while the second one is a Hopf term. As it can be
expected, the effective action stemming from (6) is identical to (18) with
$v_\mu$ (or equivalently $J_\mu$) is replaced by its dual counterpart which
has been defined by previously.\par
We expect that the occurence of self-duality within the class of (effective)
models considered in this letter may well have some interesting consequences
in more realistic models aiming to describe some of the physics of
condensed-matter systems. This is presently under study and will be reported
in a forthcoming publication.\par
{\bf{Acknowledgements:}} Three of us (Y.G, T.M, J.C.W) would like to thanks
B. Dolan for interesting discussions at the beginning of this work.\par 
\vfill\eject
{\noindent{\bf{REFERENCES}}}
\vskip 1 true cm
\item {[1]}: N. Seiberg and E. Witten, Nucl. Phys. B246 (1994) 19.\par
\item {[2]}: For a review see E. Kiritsis, Supersymmetry and Duality in Field
Theory and String Theory, preprint hep-th/9911525 (1999).\par
\item {[3]}: D. Carpentier, J. Phys. A: Math. Gen. 32 (1999) 3865 and
references therein.\par
\item {[4]}: J.L. Cardy, Nucl. Phys. B205 (1982) 17; J. L. Cardy and E.
Rabinovici. Nucl. Phys. B205 (1982) 1.\par
\item {[5]}: M. C. Diamantini, P. Sodano and C. A. Trugenberger, Nucl. Phys.
B448 (1995) 505; M. C. Diamantini, P. Sodano and C. A. Trugenberger, Nucl.
Phys. B474 (1996) 641;
for a general review see E. Fradkin, Field Theories of Condensed Matter
systems, Addison-Wesley, Redwood City (1991).\par
\item {[6]}: For a review see R. E. Prange and S. M. Girvin (eds), The
Quantum Hall Effect (1990), New York: Springer.\par
\item {[7]}: C. A. L\"utken and G. G. Ross, Phys. Rev. B45 (1992) 
11837; C. A. L\"utken and G. G. Ross, Phys. Rev. B48 (1993) 2500; A. Shapere
and F. Wilczek, Nucl. Phys. B320 (1989) 669; C. A. L\"utken, Nucl. Phys.
B396 (1993) 670; see also C. P. Burgess and C. A. L\"utken, Nucl. Phys. B500
(1997) 367; C. P. Burgess and C. A. L\"utken, Phys. Lett. B451 (1999) 365.\par
\item {[8]}: S. Kivelson, D. H. Lee and S. C. Zhang, Phys. Rev. B46 (1992)
2223.\par
\item {[9]}: Y. Georgelin and J. C. Wallet, Phys. Lett. A224 (1997) 303; Y.
Georgelin, T. Masson and J. C. Wallet, J. Phys. A.: Math. Gen.30 (1997)
5065; Y. Georgelin, T. Masson and J. C. Wallet, J. Phys. A: Math. Gen.33
(2000) 39; Y. Georgelin, T. Masson and J. C. Wallet, J. Phys. A.: Math. Gen.33
(2000) 8649.\par
\item {[10]}: B. P. Dolan, Nucl. Phys. B554 (1998) 487; B. P. Dolan, J.
Phys. A.: Math. Gen.32 (1999) L243, C. P. Burgess, R. Dib and B. P. Dolan,
Derivation of the semi circle law form the law of the corresponding states,
preprint cond-mat/9911476 (1999); B. P. Dolan, Duality in the Quantum Hall
Effect - the role of the electron spin, preprint cond-mat/0002228 (2000).\par
\item {[11}: S. J. Ree and A. Zee, Nucl. Phys. B352 (1991) 897.\par
\item {[12]}: E. Fradkin and S. Kivelson, Nucl. Phys. B474 (1996) 543.\par
\item {[13]}: see first ans second of ref. [5]; see also E. Dagotto, A.
Kocic and J. Kogut, Phys. Rev. Lett. 62 (1989) 1083.\par
\item {[14]}: J. Stern, Phys. Lett. B265 (1991) 119.\par
\item {[15]}; Y. Georgelin and J. C. Wallet, Mod. Phys. Lett. A7 (1992),
1149; Y. Georgelin and J. C. Wallet, Phys, Rev, D50 (1994) 6610.\par

\end